\documentclass[11pt]{article}

\usepackage{graphicx}
\usepackage{slashed}
\newcommand {\ga} {\ {\raise-.5ex\hbox{$\buildrel>\over\sim$}}\ }
\newcommand {\la} {\ {\raise-.5ex\hbox{$\buildrel<\over\sim$}}\ } 
\begin{document}

\title{A Model of Quark and Lepton Mixing and Mass Hierarchy}

\author{{\bf S.M. Barr and Heng-Yu Chen} \\
Department of Physics and Astronomy \\ 
and \\
Bartol Research Institute \\ University of Delaware \\
Newark, Delaware 19716 } 

\maketitle

\begin{abstract}
It is shown that an idea proposed in 1996 that relates in a qualitatively correct way the inter-family mass hierarchies of the up quarks, down quarks, charged leptons, and neutrinos, can be combined with a predictive scheme recently proposed for relating quark mixing and neutrino mixing. In the resulting model, the entire flavor structure of the
quarks and leptons is expressible in terms of two ``master matrices": a diagonal matrix that gives the inter-family mass ratios, and an off-diagonal matrix that controls all flavor mixing.
\end{abstract}

\section{Introduction}

The flavor problem has two aspects: explaining the pattern of quark and lepton mixing angles and explaining the pattern of inter-family mass hierarchies. In this paper we show that an idea proposed in 1996 \cite{BabuBarr1996} for explaining the mass hierarchies can be successfully combined with an idea proposed in 2012 \cite{BarrChen2012} for explaining the mixing. We shall refer to these as the ``BB idea" and the ``BC idea" respectively.
The two ideas are actually complementary, and by combining them a model emerges that is simpler and more explanatory than either by itself. We shall first briefly review the two ideas and then show how they can be combined. 

The BB idea was based on the observation that the inter-family mass hierarchy of the up quarks $(u, c, t)$ is stronger
than those of the down quarks $(d,s,b)$ and charged leptons $(e, \mu, \tau)$, which in
turn are stronger than that of the neutrinos. Ref. \cite{BabuBarr1996} pointed out that the strengths of these hierarchies correlate, in an $SU(5)$ framework, with the number
of fermion $10$-plets that appear in the corresponding Yukawa terms. Up quark masses come from $({\bf 10} \; {\bf 10} ) {\bf 5}_H$ terms, which have two factors of fermion $10$-plets. Down quark and charged lepton masses come from $({\bf 10} \; \overline{\bf {5}})  \overline{\bf {5}}_H$ terms, which have only one such factor. And the neutrino masses come from effective dimension-5 
$(\overline{{\bf 5}} \; \overline{{\bf 5}})  {\bf 5}_H {\bf 5}_H$ terms, which contain no such factors. 

The BB idea was that every fermion 10-plet in a Yukawa term is accompanied by 
a factor in the mass matrix of a hierarchical, diagonal matrix $H$, which one can write as $H =
{\rm diag} ( \alpha, \beta, 1)h$, where $\alpha \ll \beta \ll 1$. 
This can happen as the result of the mixing of the 10-plets in the usual three
chiral families, which we denote by ${\bf 10}^0_i + \overline{{\bf 5}}^0_i$, with ``extra"
vectorlike 10-plets, which we denote by ${\bf 10}'_i + \overline{{\bf 10}}'_i$ ($i = 1,2,3$). Let the ``underlying" Yukawa terms that give electroweak-breaking quark and lepton masses be of the form

\begin{equation}
({\bf 10}^0_i Y^u_{ij} {\bf 10}^0_j) {\bf 5}_H + ( {\bf 10}^0_i Y_{ij} 
\overline{{\bf 5}}^0_j ) \overline{{\bf 5}}_H + 
( {\bf 10}^0_i y_{ij} 
\overline{{\bf 5}}^0_j ) \overline{{\bf 45}}_H
+ ( \overline{{\bf 5}}^0_i Y^{\nu}_{ij} \overline{{\bf 5}}^0_j ) {\bf 5}_H {\bf 5}_H
/M_R.
\end{equation}

\noindent The role of the term with the ${\bf 45}_H$ of Higgs fields is to give different contributions to the mass matrices of the down quarks and charged leptons
\cite{GeorgiGlashow} and thus avoid the ``bad" predictions of minimal $SU(5)$ that $m_e = m_d$, and 
$m_{\mu} = m_s$ at the GUT scale. Suppose that ${\bf 10}^0_i$ and ${\bf 10}'_i$ mix in a family-diagonal
way to produce a light linear combination ${\bf 10}_i$ that contains Standard
Model fermions and an orthogonal linear combination ${\bf 10}^h_i$
that is superheavy. Then one can write

\begin{equation}
{\bf 10}^0_i = \cos \theta_i {\bf 10}_i + \sin \theta_i {\bf 10}^h_i.
\end{equation}

\noindent Substituting this into Eq. (1), one obtains for the effective Yukawa terms
of the Standard Model fermions

\begin{equation}
({\bf 10}_i \cos \theta_i Y^u_{ij} \cos \theta_j {\bf 10}_j) {\bf 5}_H + ( {\bf 10}_i \cos \theta_i Y_{ij} 
\overline{{\bf 5}}^0_j ) \overline{{\bf 5}}_H + 
( {\bf 10}_i \cos \theta_i y_{ij} 
\overline{{\bf 5}}^0_j ) \overline{{\bf 45}}_H
+ ( \overline{{\bf 5}}^0_i Y^{\nu}_{ij} \overline{{\bf 5}}^0_j ) {\bf 5}_H {\bf 5}_H
/M_R.
\end{equation}

\noindent Therefore, the effective quark and lepton mass terms of the Standard Model quarks and leptons can be written

\begin{equation}
\begin{array}{ccl} 
M_u & = & H \; m_u \; H, \\ 
M_d & = & H \; m_d, \\ 
M_{\ell} & = & m_{\ell} \; H, \\ 
M_{\nu} & = & m_{\nu},
\end{array}
\end{equation}

\noindent where

\begin{equation}
H = \left( \begin{array}{ccc} \cos \theta_1 & 0 & 0 \\ 
0 & \cos \theta_2 & 0 \\ 0 & 0 & \cos \theta_3 \end{array} \right)
\equiv \left(  \begin{array}{ccc} \alpha & 0 & 0 \\ 
0 & \beta & 0 \\ 0 & 0 & 1 \end{array} \right) h, 
\end{equation}

\noindent and where $(m_u)_{ij} = Y^u_{ij} v_5$, $\;\;(m_d)_{ij} = Y_{ij} v_{\overline{5}}
+ y_{ij} v_{\overline{45}}$, 
$\;\;(m_{\ell})_{ij} = Y_{ij} v_{\overline{5}}
-3 y_{ij} v_{\overline{45}}$, and $(m_{\nu})_{ij} = Y^{\nu}_{ij} (v^2_5/M_R)$.
These four ``underlying" mass matrices $m_u, m_d, m_{\ell}$, and $m_{\nu}$ are not
assumed to have any special form, and therefore for each of them one expects all the elements to be roughly of the same order.  From Eqs. (4) and (5) one has

\begin{equation}
\begin{array}{ll} 
M_u \sim \left( \begin{array}{ccc} \alpha^2 & \alpha \beta & \alpha \\
\alpha \beta & \beta^2 & \beta \\
\alpha & \beta & 1 \end{array} \right) \; \mu_u, \;\;\; & 
M_{\nu} \sim  \left( \begin{array}{ccc} 1 & 1 & 1 \\
1 & 1 & 1 \\ 1 & 1 & 1 \end{array} \right) \; \mu_{\nu} 
\\ & \\
M_d \sim \left( \begin{array}{ccc} \alpha & \alpha  & \alpha \\
\beta & \beta & \beta \\
1 & 1 & 1 \end{array} \right) \mu_d, \;\;\; &  
M_{\ell} \sim \left( \begin{array}{ccc} \alpha &  \beta & 1 \\
\alpha  & \beta & 1 \\
\alpha & \beta & 1 \end{array} \right) \; \mu_{\ell},
\end{array}
\end{equation}

\noindent where ``$\sim$" means that the various elements are of the given order 
of magnitude. This obviously gives

\begin{equation}
\begin{array}{lcl}
m_u : m_c : m_t & \sim & \alpha^2 : \beta^2 : 1 \\
m_d : m_s : m_b & \sim & \alpha : \beta : 1 \\
m_e : m_{\mu} : m_{\tau} & \sim & \alpha : \beta : 1 \\
m_{\nu_1} : m_{\nu_2} : m_{\nu_3} & \sim & 1 : 1 : 1.
\end{array}
\end{equation}

\noindent This reproduces well, in a qualitative way, the strengths of the inter-family mass hierarchies of the different types of fermions. Also from inspection of Eq. (6) it is apparent that

\begin{equation}
U_{MNS} \sim \left( \begin{array}{ccc}
1 & 1 & 1 \\ 1 & 1 & 1 \\ 1 & 1 & 1 \end{array} \right), \;\;\;
V_{CKM} \sim \left( \begin{array}{ccc}
1 & \alpha/\beta & \alpha \\
\alpha/\beta & 1 & \beta \\
\alpha & \beta & 1 \end{array} \right).
\end{equation}

\noindent This gives $O(1)$ MNS mixing angles and small CKM mixing angles, with
$|V_{ub}| \sim |V_{us} V_{cb}|$, which also is qualitatively correct.
On the other hand, since there are no constraints on the forms of the four
underlying $3 \times 3$ mass matrices $m_u$, $m_d$, $m_{\ell}$, and $m_{\nu}$, 
the BB idea in this form has many free parameters and can only make
qualitative post-dictions rather than precise quantitative predictions.

We turn now to a review of the BC idea \cite{BarrChen2012}. The BC idea 
is that all inter-family mixing among the Standard
Model fermions arises from a single source.
This source is the mixing between the $\overline{{\bf 5}}$ multiplets 
in the three chiral families and those in extra 
$\overline{{\bf 5}} + {\bf 5}$ vectorlike pairs. Because mixing comes from 
the $\overline{{\bf 5}}$ multiplets of $SU(5)$, there is large mixing only for 
left-handed leptons and right-handed quarks, thus also explaining why the MNS mixing
is large and the CKM mixing is small. (This is the basic idea of so-called 
``lopsided models" \cite{BabuBarr1996,lopsided}.) 

The specific assumption in the BC model 
is that if there existed {\it only} the three chiral families, then the 
quark and lepton mass matrices would be diagonal due to abelian
family symmetries. But the extra $\overline{{\bf 5}} + {\bf 5}$ vectorlike
multiplets,
which are assumed {\it not} to transform under the family
symmetries, are able to mix with the chiral families due
to spontaneous breaking of those symmetries.  
This induces mixing among the families of Standard Model
quarks and leptons.

In the BC model, the fermion content consists of 
$({\bf 10}^0_i + \overline{{\bf 5}}^0_i) + 
(\overline{{\bf 5}}'_A + {\bf 5}'_A)$,
where $i = 1,2,3$ and $A = 1,2, ... ,N$. There are abelian symmetries 
$Z_2^{(1)} \times Z_2^{(2)} \times Z_2^{(3)}$, such that under 
$Z_2^{(j)}$ the fields ${\bf 10}_i$ and $\overline{{\bf 5}}_i$
are odd if $i = j$ but even if $i \neq j$. As a consequence, the
``underlying" Yukawa terms of the three chiral families
have the family-diagonal form 
 
\begin{equation}
({\bf 10}^0_i Y^u_i {\bf 10}^0_i) {\bf 5}_H + ( {\bf 10}^0_i Y_i 
\overline{{\bf 5}}^0_i ) \overline{{\bf 5}}_H + 
( {\bf 10}^0_i y_i 
\overline{{\bf 5}}^0_i ) \overline{{\bf 45}}_H
+ ( \overline{{\bf 5}}^0_i Y^{\nu}_i \overline{{\bf 5}}^0_i ) {\bf 5}_H {\bf 5}_H
/M_R.
\end{equation}

The abelian family symmetries are broken spontaneously by 
the VEVs of Standard-Model-singlet scalars ${\bf 1}_{Hi}$, 
$i=1,2,3$. (Like the fermions, ${\bf 1}_{Hi}$ is odd under
$Z_2^{(j)}$ if $i=j$ and even otherwise.) This allows the 
chiral fermions and the extra vectorlike fermions to mix through
the following Yukawa terms:
$y_{iA} (\overline{{\bf 5}}^0_i {\bf 5}'_A)\langle {\bf 1}_{Hi}
\rangle + y'_{AB} (\overline{{\bf 5}}'_A {\bf 5}'_B) \langle {\bf 1}_H
\rangle$. Because of these terms, the $\overline{{\bf 5}}^0_i$ and 
$\overline{{\bf 5}}'_A$ mix to produce three linear combinations, 
$\overline{{\bf 5}}_i$, which contain Standard Model quarks and 
leptons, and $N$ linear combinations $\overline{{\bf 5}}^h_A$ that
are orthogonal to them and superheavy. One can therefore write

\begin{equation}
\overline{{\bf 5}}^0_i = A_{ij} \overline{{\bf 5}}_j
+ B_{iA} \overline{{\bf 5}}^h_A,
\end{equation}

\noindent where $A$ and $B$ are non-diagonal matrices satisfying
$A A^{\dag} + B B^{\dag} = I$. Substituting Eq. (10) into Eq. (9), one finds that
the effective mass terms for the Standard Model quarks and leptons can
be written

\begin{equation}
\begin{array}{ccl}
M_u & = & m_u, \\ 
M_d & = & m_d \; A,  \\ 
M_{\ell} & = & A^T \; m_{\ell}, \\ 
M_{\nu} & = & A^T \; m_{\nu} \; A,
\end{array}
\end{equation}

\noindent where $(m_u)_{ij} = \delta_{ij} Y^u_i v_5$, $\;\;(m_d)_{ij} = 
\delta_{ij}(Y_i v_{\overline{5}} + y_i v_{\overline{45}})$, 
$\;\;(m_{\ell})_{ij} = \delta_{ij} (Y_i v_{\overline{5}}
-3 y_i v_{\overline{45}})$, and $(m_{\nu})_{ij} = \delta_{ij} Y^{\nu}_i (v^2_5/M_R)$.
The non-zero elements of these diagonal matrices are free parameters of the
model. To fit the observed quark and lepton masses they must be hierarchical, but the 
BC model does not attempt to explain these hierarchies or relate them to each other.

As shown in \cite{BarrChen2012}, the matrix $A$ can be brought by changes of basis to 
the form

\begin{equation}
A = {\cal D} A_{\Delta} {\cal U},
\end{equation}

\noindent where ${\cal U}$ is a unitary matrix, ${\cal D}$ is the diagonal matrix ${\rm diag}
(\delta, \epsilon, 1) d$, and $A_{\Delta}$ is a triangular matrix of the form

\begin{equation}
A_{\Delta} = \left( \begin{array}{ccc} 
1 & b & c e^{i \theta} \\ 0 & 1 & a \\ 0 & 0 & 1 \end{array}
\right),
\end{equation}

\noindent and $a$, $b$, and $c$ are real numbers. The matrix ${\cal U}$ 
can be eliminated by redefining the fields of the right-handed quarks and left-handed leptons. These redefinitions do not affect the CKM or MNS mixing matrices.
(They do not affect the CKM matrix, as only a redefinition of the {\it right-handed}
quarks is involved. They do not affect the MNS mixing matrix, since the same
redefinition is done on the neutrinos and charged leptons.) The diagonal matrix
${\cal D}$ can be absorbed into the diagonal matrices: $\overline{m}_u \equiv
m_u$, $\overline{m}_d \equiv m_d {\cal D}$, $\overline{m}_{\ell} \equiv
m_{\ell} {\cal D}$, and $\overline{m}_{\nu} = m_{\nu} {\cal D}^2$. After these
redefinitions one has

\begin{equation}
\begin{array}{ccl}
M_u & = & \overline{m}_u, \\ 
M_d & = & \overline{m}_d \; A_{\Delta},  \\ 
M_{\ell} & = & A_{\Delta}^T \; \overline{m}_{\ell}, \\ 
M_{\nu} & = & A_{\Delta}^T \; \overline{m}_{\nu} \; A_{\Delta}.
\end{array}
\end{equation}

\noindent The elements of the diagonal matrix $\overline{m}_u$ are obviously just
the masses of the up quarks, $\overline{m}_u = {\rm diag} (m_u, m_c, m_t)$. 
One can also easily show that to a very good approximation the elements of the diagonal matrices $\overline{m}_d$ and
$\overline{m}_{\ell}$ are the masses of down quarks and charged leptons:
$\overline{m}_d \cong {\rm diag} (m_d, m_s, m_b)$, and 
$\overline{m}_{\ell} \cong {\rm diag} (m_e, m_{\mu}, m_{\tau})$. That 
means that in the basis in which the up quark mass matrix is diagonal
the mass matrix of the down quarks is given by

\begin{equation}
M_d = \left( \begin{array}{ccc}
m_d & m_d b & m_d c e^{i \theta} \\
0 & m_s & m_s a \\
0 & 0 & m_b \end{array} \right).
\end{equation}

\noindent  From this one can read off directly that $|V_{us}| \cong
\frac{m_d}{m_s} b$, $|V_{cb}| \cong \frac{m_s}{m_b} a$, and $V_{ub}
\cong \frac{m_d}{m_b} c e^{i \theta}$. Therefore, the triangular matrix
$A_{\Delta}$ can be written

\begin{equation}
A_{\Delta} = \left( \begin{array}{ccc} 
1 & \left| \frac{m_s}{m_d} V_{us} \right| & 
\left| \frac{m_b}{m_d} V_{ub} \right| e^{i \delta_{KM}} \\
0 & 1 & \left| \frac{m_b}{m_s} V_{cb} \right| \\
0 & 0 & 1 \end{array} \right).
\end{equation}

The mass matrix $M_{\ell} = A_{\Delta}^T \overline{m}_{\ell}$ 
of the charged leptons, given in Eq. (14), is obviously not diagonal.
However, it is easily seen that it is diagonalized by rotations
of the {\it right-handed} leptons by angles that are of order $m_e/m_{\mu}$, $m_{\mu}/m_{\tau}$, and $m_e/m_{\tau}$, while the required rotations
of the {\it left-handed} leptons are only of order the squares of
these ratios, and thus negligible. As far as the left-handed charged
leptons are concerned, therefore, we are effectively in the mass basis.
Combining Eq. (14) and (16), one has the following expression for 
the neutrino mass matrix 

\begin{equation}
M_{\nu} = \mu_{\nu} \left( \begin{array}{ccc}
1 & 0 & 0 \\
\left| \frac{m_s}{m_d} V_{us} \right| & 1 & 0 \\
\left| \frac{m_b}{m_d} V_{ub} \right| e^{i \delta_{KM}} & \left| \frac{m_b}{m_s} V_{cb} \right|
& 1 \end{array} \right)
\left( \begin{array}{ccc} 
q e^{i \theta_q} & 0 & 0 \\
0 & p e^{i \theta_p} & 0 \\
0 & 0 & 1 \end{array} \right)
\left( \begin{array}{ccc}
1 & \left| \frac{m_s}{m_d} V_{us} \right| & \left| 
\frac{m_b}{m_d} V_{ub} \right| e^{i \delta_{KM}} \\
0 & 1 & \left| \frac{m_b}{m_s} V_{cb} \right| \\
0 & 0 & 1 \end{array} \right).
\end{equation}

\noindent The five free parameters $p$, $\theta_p$, $q$, $\theta_q$,
and $\mu_{\nu}$ determine all nine neutrino properties:
the three neutrino massses, three MNS mixing angles, the Dirac
CP phase of the leptons, and the two Majorana CP phases. So 
there are four predictions. These are discussed in detail in \cite{BarrChen2012}.
What was found there was that a good fit is obtained to the three neutrino
mixing angles and to the neutrino mass splittings, and that the Dirac CP phase
of the leptons is predicted to be roughly 210 degrees. 
The values of the parameters that gave the best fits were
$p = 0.1525$, $q = 0.01405$, $\theta_p = -2.73$ radians, and $\theta_q = -0.352$ radians.
The values of $p$ and $q$ are important for our later discussions.  

\section{Combining the two ideas}

In the BB idea, all the inter-family mass hierarchies come from 
the single matrix $H$, while in the BC idea all the inter-family mixing 
comes from the single matrix $A$. The question naturally arises whether
these two ideas can be combined in such a way that the whole flavor structure
can be accounted for with {\it only} the matrices $A$ and $H$,
thereby producing a more predictive and explanatory model. The answer is yes,
as we shall now show by describing a specific model that does this.

The fermion content of the model consists of the following $SU(5)$
multiplets:

\begin{equation}
({\bf 10}^0_i + \overline{{\bf 5}}^0_i)_{i = 1,2,3} 
\;\; + \;\; ({\bf 10}'_A + \overline{{\bf 10}}'_A)_{A = 1,2,3} 
\;\; + \;\; (\overline{{\bf 5}}'_m  + {\bf 5}'_m)_{m = 1, 2, ... N}.
\end{equation}

\noindent Yukawa terms involving only ${\bf 10}^0_i$ and 
$\overline{{\bf 5}}^0_i$ will give rise to ``underlying" mass
matrices that get multiplied by factors of the matrices
$H$ and $A$. In order for $H$ and $A$ to account for
{\it all} the flavor structure, the underlying mass matrices
should have a trivial flavor structure, {\it i.e.} they should 
be proportional to the identity matrix.. This can be the
case if there is an $SO(3)$ family symmetry under which
the ${\bf 10}^0_i$ and $\overline{{\bf 5}}^0_i$ transform
as triplets. The underlying Yukawa terms would then have the form

\begin{equation}
Y_u ({\bf 10}^0_i \; {\bf 10}^0_i) \; {\bf 5}_H \;\; + \;\; 
Y_d ({\bf 10}^0_i  \; \overline{{\bf 5}}^0_i) \; \overline{{\bf 5}}_H
\;\; + \;\; Y_{\nu} (\overline{{\bf 5}}^0_i \; \overline{{\bf 5}}^0_i) 
\; {\bf 5}_H {\bf 5}_H/M_R.
\end{equation}

\noindent Note that unlike Eqs. (1) and (9) there is no term here
with the $\overline{{\bf 45}}_H$ of Higgs fields. Since all
the underlying Yukawa terms must be flavor-independent, due to the $SO(3)$ symmetry,
adding a term with the $\overline{{\bf 45}}_H$ in Eq. (19)
would still leave the down quark and charged
lepton mass matrices proportional to each other at the GUT scale.
Therefore, the group-theoretical factors needed to avoid the ``bad" 
minimal $SU(5)$ relation $M_d = M_{\ell}^T$ must appear in either the
$H$ or $A$ matrices. In the model we are describing, they will appear in
the $H$ matrix, as will be seen. 

The matrix $A$ arises, in exactly the manner explained earlier, 
from the mixing of the 
$\overline{{\bf 5}}^0_i$ with the ``extra" $\overline{{\bf 5}}'_m$,
which are assumed not to transform under any flavor symmetry. 
Let there be
at least two Standard-Models-singlet Higgs fields that are triplets
under $SO(3)$, denoted by ${\bf 1}_H^{ni}$, where $n$ labels
the Higgs triplet and $i$ is the $SO(3)$ index. Then one can write 
the following mass and Yukawa terms for the fermion 5-plets:

\begin{equation}
\begin{array}{l}
M_{mn} ({\bf 5}'_m \; \overline{{\bf 5}}'_n ) \; + \; 
y_{mn} ({\bf 5}'_m \; \overline{{\bf 5}}^0_i) \; \langle 
{\bf 1}_H^{ni} \rangle  \\ \\
= {\bf5}'_m (M_{mn} \overline{{\bf 5}}'_m \; + \; \Delta_{mi} 
\overline{{\bf 5}}^0_i),
\end{array}
\end{equation}

\noindent where $\Delta_{mi} = \sum_n y_{mn} \langle {\bf 1}_H^{ni}
\rangle$. We assume that the matrices $M$ and $\Delta$ are
superheavy and of the same order. (For example, they
may both be of order the GUT scale.) These terms will make $N$ 
linear combinations of the $\overline{{\bf 5}}$ fields superheavy
and leave three linear combinations light. These light
linear combinations, which contain Standard Model quarks and leptons,
will be denoted $\overline{{\bf 5}}_i$. The superheavy combinations 
will be denoted by $\overline{{\bf 5}}^h_m$.

It is easily seen that if $A \equiv [I + T^{\dag} T]^{-1/2}$
and $B \equiv [I + T^{\dag} T]^{-1/2} T^{\dag}$, where 
$T = M^{-1} \Delta$, then
$\overline{{\bf 5}}^0 = A \overline{{\bf 5}} + B \overline{{\bf 5}}^h$.
Exactly as in the BC model, when substituted into Eq. (19),
this leads to factors of $A$ in the effective mass matrices of the 
Standard Model quarks and leptons.

The factors of $H$ in those matrices arise, as in the BB scheme, from the
mixing of ${\bf 10}^0_i$ with the ${\bf 10}'_A$. In order for $H$ to
come out diagonal, the ${\bf 10}'_A + \overline{{\bf 10}}'_A$ must 
transform under a flavor symmetry. A simple possibility is
an abelian symmetry $Z_2^{(1)} \times Z_2^{(2)} \times Z_2^{(3)}$,
such that ${\bf 10}'_A$ and $\overline{{\bf 10}}'_A$ are odd under 
$Z_2^{(B)}$ if $A = B$ and even otherwise. Let there be 
three Standard-Model-singlet Higgs fields ${\bf 1}_H^{Ai}$, which
are triplets under $SO(3)$ and transform under 
$Z_2^{(1)} \times Z_2^{(2)} \times Z_2^{(3)}$ in the obvious way. Then 
the following mass and Yukawa terms of the fermion 10-plets 
are allowed

\begin{equation}
\overline{{\bf 10}}'_A \left( Y_A {\bf 1}_H + y_A {\bf 24}_H \right) {\bf 10}'_A 
\; + \; \overline{{\bf 10}}'_A \left( Y'_A {\bf 1}^{Ai}_H + y'_A {\bf 24}^{Ai}_H \right) {\bf 10}^0_i. 
\end{equation}

\noindent The role of the adjoint Higgs fields ${\bf 24}_H$ and ${\bf 24}^{Ai}_H$ is
to introduce $SU(5)$ breaking into the quark and lepton
mass matrices, through $H$, and thus avoid the ``bad"
minimal $SU(5)$ prediction that the down quark masses equal 
the charged lepton masses at the GUT scale. It is notationally 
simpler, however, to explain the mixing of the 10-plets without considering
the effects of the adjoint fields in Eq. (21), 
so we will first discuss the unrealistic case where their
VEVs are set to zero (which we will call the ``minimal model")
and then later discuss the realistic case where
their VEVs are non-zero. 

If certain coefficients in the Higgs potential are positive
then the VEVs of ${\bf 1}_H^{Ai}$ will be orthogonal to each
other in $SO(3)$ space: $\sum_i \langle {\bf 1}_H^{Ai} \rangle
\langle {\bf 1}_H^{Bi} \rangle = c_A \delta_{AB}$. (In particular, if
the coefficients of the terms $\left( \sum_{i=1}^3 {\bf 1}_H^{Ai}
{\bf 1}_H^{Bi} \right)^2$ are positive it will ensure this orthogonality.)
Without loss of generality, one can then choose a basis in $SO(3)$ space 
such that the axes are aligned with the VEVs of the three
singlet VEVs. That is, so that
$\langle {\bf 1}_H^{Ai} \rangle = s_A \delta^{Ai}$. 
Defining, $Y_A \langle {\bf 1}_H \rangle \equiv M_A$ and
$Y'_A \langle {\bf 1}^{Ai}_H \rangle \equiv \Delta_A \delta^{Ai}$, Eq. (21) with
adjoint VEVs set to zero gives

\begin{equation}
\overline{{\bf 10}}'_A \left( M_A {\bf 10}'_A \; + \; \Delta_A \delta^{Ai} {\bf 10}^0_i \right).
\end{equation}

\noindent The three linear combinations of 10-plets appearing with the parentheses
in Eq. (22) are superheavy and will be denoted ${\bf 10}^h_A$, whereas
the three linear combinations $(
- \Delta_A {\bf 10}'_A \; + \; M_A \delta^{Ai} {\bf 10}^0_i)$ that are orthogonal to them contain Standard Model fermions and will be denoted ${\bf 10}_i$. This gives

\begin{equation}
{\bf 10}^0_i = \cos \theta_i  \; {\bf 10}_i  
\; + \; \sin \theta_i \; \delta^{Ai} \; {\bf 10}^h_A, 
\end{equation}

\noindent where $\cos \theta_i \equiv \delta^{Ai} M_A/\sqrt{|M_A|^2 + |\Delta_A|^2}$ and 
$\sin \theta_i \equiv \delta^{Ai} \; \Delta_A/\sqrt{|M_A|^2 + |\Delta_A|^2}$. 
Substituting this into Eq. (19), one finds that every factor of ${\bf 10}_i$ in
the effective Yukawa couplings of the Standard Model fermions is accompanied by a factor
of $\cos \theta_i$, as in Eq. (3).
We will assume a
hierarchical pattern $|\Delta_1/M_1| \gg |\Delta_2/M_2| \gg 1 \gg |\Delta_3/ M_3|$. 
Then we can define a matrix $H$ by 

\begin{equation}
H \equiv \left( \begin{array}{ccc} \cos \theta_1 & 0 & 0 \\ 
0 & \cos \theta_2 & 0 \\ 0 & 0 & \cos \theta_3 \end{array} \right)
\equiv \left(  \begin{array}{ccc} \alpha & 0 & 0 \\ 
0 & \beta & 0 \\ 0 & 0 & \gamma \end{array} \right), \;\; {\rm where} \; \alpha \ll \beta \ll \gamma \cong 1. 
\end{equation}

\noindent Substituting $\overline{{\bf 5}}^0 = A \overline{{\bf 5}} + B \overline{{\bf 5}}^h$ and Eq. (23) into Eq. (19) and using Eq. (24), the effective mass matrices
of the Standard Model quarks and leptons can then be written

\begin{equation}
\begin{array}{lcl}
M_u = (H^2) \; \mu_u & & \\ 
M_d = (H \; A) \; \mu_d & \longrightarrow &
M_d = (H \; {\cal D}) \; A_{\Delta} \; \mu_d, \\
M_{\ell} = (A^T \; H) \; \mu_d & \longrightarrow &
M_{\ell} = A_{\Delta}^T \; ( {\cal D} \; H) \; \mu_d, \\
M_{\nu} = (A^T \; A) \; \mu_{\nu} & \longrightarrow &
M_{\nu} = A_{\Delta}^T \; ({\cal D}^2) \; A_{\Delta} \; \mu_{\nu}, 
\end{array}
\end{equation}

\noindent This is the basic result of the model. Other than certain overall mass scales
($\mu_u$, $\mu_d$, and $\mu_{\nu}$) all the flavor structure of the quarks and leptons is controlled by two matrices: a mixing matrix $A$ and a hierarchy matrix $H$. In going to the last expressions in each line of Eq. (25), we have used 
$A = {\cal D} A_{\Delta} {\cal U}$
and absorbed ${\cal U}$ by field redefinitions (as explained
previously). We write the matrix
${\cal D}$ as ${\cal D} = {\rm diag} ( \delta, \epsilon, 1)d$ and absorb the factors 
of $d$ into redefined mass scales $\mu'_d$ and $\mu'_{\nu}$. One therefore ends up with
the following result (for the ``minimal" version of the model):

\begin{equation}
\begin{array}{ccl}
M_u & = & \left(  \begin{array}{ccc} 
|\alpha|^2 & 0 & 0 \\
0 & |\beta|^2 & 0 \\
0 & 0 & 1 \end{array} \right) \; \mu_u, \\ & & \\
M_d & = & M_{\ell}^T \; = \; \left( \begin{array}{ccc} 
|\alpha \delta| & 0 & 0 \\
0 & |\beta \epsilon | & 0 \\
0 & 0 & 1 \end{array} \right) \left( \begin{array}{ccc}
1 & b & c e^{i \theta} \\
0 & 1 & a \\
0 & 0 & 1 \end{array} \right) \; \mu'_d, \\ & & \\
M_{\nu} & = & 
\left( \begin{array}{ccc}
1 & 0 & 0 \\
b & 1 & 0 \\
c e^{i \theta} & a & 1 \end{array} \right) 
\left( \begin{array}{ccc}
\delta^2 & 0 & 0 \\
0 & \epsilon^2 & 0 \\
0 & 0 & 1 \end{array} \right) 
\left( \begin{array}{ccc}
1 & b & c e^{i \theta} \\
0 & 1 & a \\
0 & 0 & 1 \end{array} \right) \; \mu'_{\nu}.
\end{array}
\end{equation}

\noindent  Of course, the form obtained for $M_{\nu}$ is the same as shown in
Eq. (17). The parameters called $pe^{i \theta_p}$ and $q e^{i \theta_q}$ in Eq. (17) are here called $\epsilon^2$ and $\delta^2$. It should be noted that in Eq. (26), the phases of $\delta$, $\epsilon$, $\alpha$, and $\beta$ do not matter for the matrices $M_u$, $M_d$, and $M_{\ell}$, as they can be absorbed into the fermion fields.
But for the neutrino mass matrix $M_{\nu}$ the phases of $\delta$
and $\epsilon$ do make a difference, and have to take definite values to fit the neutrino masses and maxing angles. 

One easily sees from Eq. (26) that in this ``minimal model" one has, to very good approximation, the following ``postdictions":

\begin{equation}  
\begin{array}{rcl}
m_u : m_c : m_t & = & |\alpha|^2 : |\beta|^2 : 1, \\ & & \\
m_d : m_s : m_b \; = \; m_e : m_{\mu} : m_{\tau} & = & 
|\alpha \delta| : |\beta \epsilon| : 1, \\ & & \\
q^2 : p^2 : 1 & = & |\delta|^2 : |\epsilon|^2 : 1.
\end{array}
\end{equation}

\noindent 
From fitting the neutrino masses and mixing angles \cite{BarrChen2012}, one can determine $|\epsilon| = \sqrt{p} \cong \sqrt{0.1525} = \frac{1}{2.56}$ and $|\delta| = \sqrt{q} \cong \sqrt{0.0141} = \frac{1}{8.44}$. (See the discussion
after Eq. (17).) And one can obtain the values of $|\alpha|$ and $|\beta|$ directly from the up quark mass
ratios: $|\beta| = \sqrt{m_c/m_t} = \frac{1}{17.8}$ and $|\alpha| = \sqrt{m_u/m_t} = \frac{1}{393}$. (We take the fermion masses here and in the following equation to be
the masses at $2 \times 10^{16}$ as
run up to that scale using the Standard Model renormalizaton group equations \cite{Bora}.)
From these values one has the following result:

\begin{equation}
\begin{array}{lccc}
{\rm minimal \; model \; hierarchy} &  |\alpha \delta| : |\beta \epsilon| : 1 & = & \frac{1}{3,317} : \frac{1}{45.6} : 1 \\ & & & \\  
{\rm actual \; lepton \; ratios} & m_e : m_{\mu} : m_{\tau} & = & \frac{1}{3,650} : \frac{1}{17.3} : 1 \\ & & & \\
{\rm actual \; quark \; ratios} & m_d : m_s : m_b & = & \frac{1}{865} : \frac{1}{45.6} : 1.
\end{array}
\end{equation}

\noindent One sees that the minimal model works surprisingly well, in fact better
than in the BB idea taken by itself, where 
the inter-family mass ratios of the charged leptons and of the down quarks   
are $\alpha : \beta : 1$, as shown in Eq. (7). (That would give
$m_e/m_{\tau} \sim m_d/m_b \sim \alpha \sim \frac{1}{393}$, which is off by
an order of magnitude for the electron.) Thus the factors of $\delta$ and $\epsilon$,
which come from combining the BB and BC ideas, give more realistic down quark and charged lepton mass hierarchies. 

The combined model we are describing (so far in a minimal form) is more explanatory than the BC model. In the
BC model the inter-family mass hierarchies of the up quarks, down quarks, charged
leptons, and neutrinos are completely unrelated, being determined by
four diagonal matrices whose elements are free parameters. In the combined model,
these hierarchies are all related, and related in a way that we have just seen is qualitatively correct.
The 12 parameters in the four hierarchical diagonal matrices of the BC model are replaced
by just 7 parameters in the minimal model: $|\alpha|$, $|\beta|$, $|\delta|$, $|\epsilon|$, $\mu_u$, 
$\mu'_d$, and $\mu'_{\nu}$. This would be a huge increase in predictivity,
but of course it is {\it too} predictive, since the minimal model gives
the ``bad" minimal $SU(5)$ prediction that the charged lepton masses are equal to
the down quark masses at the GUT scale. To cure this problem requires that group-theoretical factors reflecting the breaking of $SU(5)$ appear in the fermion mass matrices. The simplest way
for this to happen is through the matrix $H$ as a result of the adjoint Higgs fields
in Eq. (21) getting non-zero VEVs. We shall now look at this in detail.

\section{The group-theoretical factors that distinguish $M_d$ from $M_{\ell}$}

As can be seen from Eq. (28), the group-theoretical factors must enhance the muon mass and the $d$ quark mass by about a factor of about 3 while having little effect on the other quark and lepton masses. One obtains a similar conclusion if one runs the fermion masses assuming low-energy supersymmetry. Using the results of \cite{Bora}, where the masses are run up to $2 \times 10^{16}$ GeV, in the MSSM with $\tan \beta = 10$, one finds

\begin{equation}
\begin{array}{llcl}
& \left(m_u, m_c, m_t\right)/m_t & = & \left( \frac{1}{179,500}, \frac{1}{368}, 1 \right),
\\ & & & \\ \Longrightarrow & (\alpha, \beta, 1) \equiv  
\left( \sqrt{\frac{m_u}{m_t}}, \sqrt{\frac{m_c}{m_t}}, 1 \right) & = &
\left( \frac{1}{424}, \frac{1}{19.2}, 1 \right),
\\ & & & \\
\Longrightarrow & (|\alpha \delta|, |\beta \epsilon|, 1) & = & \left( \frac{1}{3,575} : \frac{1}{49.1} : 1 \right), 
\\ & & & \\
& \left( m_d, m_s, m_b \right)/m_b & = & \left( \frac{1}{1142}, \frac{1}{60.14}, 1 \right), 
\\ & & & \\
& \left( m_e, m_{\mu}, m_{\tau} \right)/m_b & = & \left(
\frac{1}{2,967}, \frac{1}{14.1}, 1.24 \right). 
\end{array}
\end{equation}

\noindent These imply that 
\begin{equation}
\begin{array}{lcc}
\left( \frac{m_d}{|\alpha \delta|}, \frac{m_s}{| \beta \epsilon|},  m_b\right)/m_b
& = & \left( 3.13, 0.817, 1 \right), 
\\ & &  \\
\left( \frac{m_e}{|\alpha \delta|}, \frac{m_{\mu}}{| \beta \epsilon|}, 
m_{\tau}\right)/m_b
& = & \left( 1.21, 3.49, 1.24 \right).  
\end{array}
\end{equation}

\noindent The ratios given in Eq. (30), which are all predicted to be equal to 1 in the minimal model, must be accounted for by the group-theoretical factors. 

Seemingly, the simplest way to do this is through the coupling of adjoint Higgs fields to the 10-plets of fermions, as shown in Eq. (21). Let us first just consider the effect of the VEV of the ${\bf 24}_H$, which couples as $\overline{{\bf 10}}'_A (y_A {\bf 24}_H) 
{\bf 10}'_A$. If we define $\kappa_A$ by $\frac{y_A \langle {\bf 24}_H \rangle}{Y_A
\langle {\bf 1}_H \rangle} = \kappa_A Y_f/2$, where $f$ stands for the fermion type
$u, u^c, d$, or $\ell^c$, then the effect is that in Eq. (22), 
$M_A$ gets replaced by $M_A(1 + \kappa_A Y_f/2)$. Suppose that we assume that
$|\Delta_1/M_1| \gg |\Delta_2/M_2| \gg |\Delta_3/M_3| \sim 1$, then the angles defined after Eq. (23) are different for different fermion types and given approximately by 

\begin{equation}
\begin{array}{rcl}
\cos \theta^f_1 & \cong & \left| \frac{M_1}{\Delta_1} \left( 1 + \kappa_1 Y_f/2 \right) \right| \\ & & \\
\ll \cos \theta^f_2 & \cong & \left| \frac{M_2}{\Delta_2} \left( 1 + \kappa_2 Y_f/2 \right) \right| \\ & & \\
\ll \cos \theta^f_3 & \cong & \left[ 1 + \left| \frac{\Delta_3}{M_3} \right|^2 
(1 + \kappa_3 Y_f/2)^{-2} \right|^{-1/2} \cong 1,
\end{array}
\end{equation}

\noindent where $Y_f/2$ is the weak hypercharge of the fermion of type $f$. Then the matrix $H$ defined in Eq. (24) is replaced by matrices $H_f$, which are different for different types of fermion in the 10-plets:

\begin{equation}
H_f \equiv \left( \begin{array}{ccc} \cos \theta^f_1 & 0 & 0 \\ 
0 & \cos \theta^f_2 & 0 \\ 0 & 0 & \cos \theta^f_3 \end{array} \right), 
\end{equation}

\noindent and the fermion mass matrices have the forms

\begin{equation}
\begin{array}{lcl}
M_u = (H_u H_{u^c}) \; \mu_u & & \\ 
M_d = (H_d \; A) \; \mu_d & \longrightarrow &
M_d = (H_d \; {\cal D}) \; A_{\Delta} \; \mu_d, \\
M_{\ell} = (A^T \; H_{\ell^c}) \; \mu_d & \longrightarrow &
M_{\ell} = A_{\Delta}^T \; ( {\cal D} \; H_{\ell^c}) \; \mu_d, \\
M_{\nu} = (A^T \; A) \; \mu_{\nu} & \longrightarrow &
M_{\nu} = A_{\Delta}^T \; ({\cal D}^2) \; A_{\Delta} \; \mu_{\nu}, 
\end{array}
\end{equation}
 
If we consider the masses of the charged fermions of the second and third families, there are four mass ratios $(\frac{m_c}{m_t}, \frac{m_s}{m_b}, \frac{m_{\mu}}{m_{\tau}}$, and $\frac{m_{\tau}}{m_b}$) that must be fit using the parameters in Eq.(31), and there are four
such parameters, namely $|\Delta_3/M_3|, \kappa_3,
|\Delta_2/M_2|$, and $\kappa_2$. 

Consider first the ratio $m_{\tau}/m_b$. As is well-known this is predicted in minimal $SU(5)$ to be 1 at the GUT scale, as is also the case in the minimal version of the  
present model. In reality, however, this ratio is not exactly 1, though it is near to 1 (especially in the MSSM). In fact, for $\tan \beta = 10$ it is 1.24 at the GUT scale as shown in Eq. (30). With
the group-theoretic factors of Eq. (31) one sees that it is given by

\begin{equation}
1.24 = \left( \frac{m_{\tau}}{m_b} \right)_{M_{GUT}}=
\frac{\cos \theta^{\ell^c}_3}{\cos \theta^d_3} 
= \sqrt{\frac{ 1 + \left| \frac{\Delta_3}{M_3} \right|^2 \left( 1 + \frac{1}{6} \kappa_3
\right)^{-2}}{ 1 + \left| \frac{\Delta_3}{M_3} \right|^2 \left( 1 + \kappa_3
\right)^{-2}}},
\end{equation}

\noindent which is indeed close to but not exactly 1, for $\Delta_3/M_3 < 1$. We can also write (putting in the values given in Eq. (30)):

\begin{equation}
0.817 = \frac{m_s/m_b}{\epsilon \sqrt{m_c/m_t}} = 
\sqrt{\frac{1 + \frac{1}{6} \kappa_2}{1 - \frac{2}{3} \kappa_2}}
\left( \frac{1 + | \frac{\Delta_3}{M_3}|^2 (1 + \frac{1}{6} \kappa_3)^{-2}}
{1 + | \frac{\Delta_3}{M_3}|^2 (1 - \frac{2}{3} \kappa_3)^{-2}}
   \right)^{1/4},
\end{equation}

\noindent and

\begin{equation}
3.49 = \frac{m_{\mu}/m_b}{\epsilon \sqrt{m_c/m_t}} = 
\frac{1 + \kappa_2}{\sqrt{(1 + \frac{1}{6} \kappa_2)(1 - \frac{2}{3} \kappa_2)}}
\left( \frac{1 + | \frac{\Delta_3}{M_3}|^2 (1 + \frac{1}{6} \kappa_3)^{-2}}
{1 + | \frac{\Delta_3}{M_3}|^2 (1 - \frac{2}{3} \kappa_3)^{-2}}
   \right)^{1/4}.
\end{equation}

\noindent Eqs. (34) to (36) contain three equations with three unknowns
$\kappa_2$, $\kappa_3$, and $|\Delta_3/M_3|$. They are solved by
the values $\kappa_2 = 11.2$, $\kappa_3 = -2$, and $|\Delta_3/M_3| = 0.86$.
The remaining ratio $m_c/m_t$ can then be fit by the choice 
$|\Delta_2/M_2| = 110$.

Fitting the first family masses is more involved. There are three additional masses
to be fit ($m_e$, $m_d$, and $m_u$), but the expressions in Eqs. (31) have only
two additional parameters ($\kappa_1$ and $|\Delta_1/M_1|$. Indeed, it turns out that there is no fit. It is for this reason that one must include the effect of
the term containing ${\bf 24}^{Ai}_H$ in Eq. (21). Actually, only one such adjoint
Higgs field is required to obtain a good fit, namely ${\bf 24}^{1i}_H$. However,
as the expressions are somewhat complicated looking, we do not show them. 

One sees, then, that introducing the group-theoretic factors required to
break the well-known minimal $SU(5)$ mass degeneracies means that the model
ends up with as many free parameters as there are in the BC model of \cite{BarrChen2012}. Thus combining that model with the BB idea leads to no increase in the number of
precise quantitative predictions. However, there is a gain in explanatoru power, in that the inter-family mass hierarchies of the different types of fermions are related
to each other in a way that is qualitative correct.

\section{The typical values of $\delta$ and $\epsilon$}

We now turn to a discussion of the values of $\delta$ and $\epsilon$, the
elements of the diagonal matrix 
${\cal D}$.
It is a non-trivial condition for the viability of the model that the same values of $|\delta|$ and $|\epsilon|$ give realistic results both for the neutrino properties and for the mass hierarchies of the down quarks and charged leptons. As we have seen, the model clears this hurdle. The fit to the neutrino properties obtained in \cite{BarrChen2012} gives $|\delta| \cong \frac{1}{8.44}$
and $|\epsilon| \cong \frac{1}{2.56}$, and these values also give realistic mass hierarchies, as shown in Eq. (27).

The question arises whether these are natural values for $|\delta|$ and $|\epsilon|$
to have. Why should there be any hierarchy in the elements of ${\cal D}$? And why should that hierarchy be parallel to the hierarchy in $H$, with the diagonal elements increasing
from the first to the third family? And why should they have these particular values?
It turns out that the values of $|\delta|$ and $|\epsilon|$ needed for good fits
are indeed natural, in the sense that they lie in the middle of the range of values that are most ``likely" given the values of the elements of the triangular matrix $A_{\Delta}$, as we will now show.

The matrix ${\cal D} = {\rm diag} (\delta, \epsilon, 1)$ arises from bringing the matrix $A$ to the form $A = {\cal D} A_{\Delta} {\cal U}$, as previously explained. The matrix $A$, in turn, is defined by $A \equiv (I + T^{\dag} T)^{-1/2}$, where $T = M^{-1} \Delta$, and $M$ and $\Delta$ are the matrices appearing in Eq. (20). It is natural to assume that the matrices $M$ and $\Delta$ are both roughly of order the grand unification scale, but there is no symmetry reason why $M$ and $\Delta$ should have any special form. Consequently, the matrix $T$ has no reason to have any special form either. 

Suppose that the elements of $T$ are treated as random complex variables all of which have the same probability distribution. For each choice of $T$, one can compute the matrix $A$, and from that determine the matrices ${\cal D}$ and $A_{\Delta}$. Not surprisingly, one
finds that the elements of ${\cal D}$ are correlated with those of $A_{\Delta}$. In fact,
simple arguments show that if the elements of $A_{\Delta}$ that we have called $a$ and $b$ are large, then typically $|\delta| \sim 1/ab$ and $|\epsilon| \sim 1/a$. Since fitting the CKM angles gives $a \sim 2$ and $b \sim 4$, as can be seen from Eq. (16), the most likely values are $|\delta| \sim 1/8$ and $|\epsilon| \sim 1/2$. 

This is confirmed by a numerical search treating the elements of $T$ as random variables. We have randomly generated one million matrices $T$ whose elements are given by
$T_{ij} = 10^{r_{ij}} e^{i \theta_{ij}}$, with $-1 < r_{ij} < +1$ and $0 < \theta_{ij} < 2 \pi$ with uniform probability distribution. We compute the matrices $A_{\Delta}$ and
${\cal D}$ for each randomly generated $T$, and require that the parameters in $A_{\Delta}$ ({\it i.e.} $a$, $b$, $c$, and $\theta$) agree with the values in Eq. (16) within experimental limits. For those that meet this requirement, we plot the values of $|\epsilon|^{-1}$ and $|\delta|^{-1}$ in Fig. 1. One sees that there indeed 
tends to be a mild hierarchy $|\delta| < |\epsilon| < 1$. The dark cross in Fig. 1 represents the values that give the best fit to the neutrino properties according to
\cite{BarrChen2012}: $\left( |\epsilon|^{-1}, |\delta|^{-1} \right) = (2.56, 8.44)$.
It is apparent from Fig. 1 that these lie in the most probable range.

\noindent {\bf Fig. 1} The values of $\left( |\epsilon|^{-1}, |\delta|^{-1} \right)$
that come from randomly generated matrices $T$ that give realistic
$A_{\Delta}$. The dark cross represents the values that give the best fit
to neutrino properties: $\left( |\epsilon|^{-1}, |\delta|^{-1} \right) = (2.56, 8.44)$.

\begin{figure}[h]
\begin{center}
\includegraphics[width=15cm]{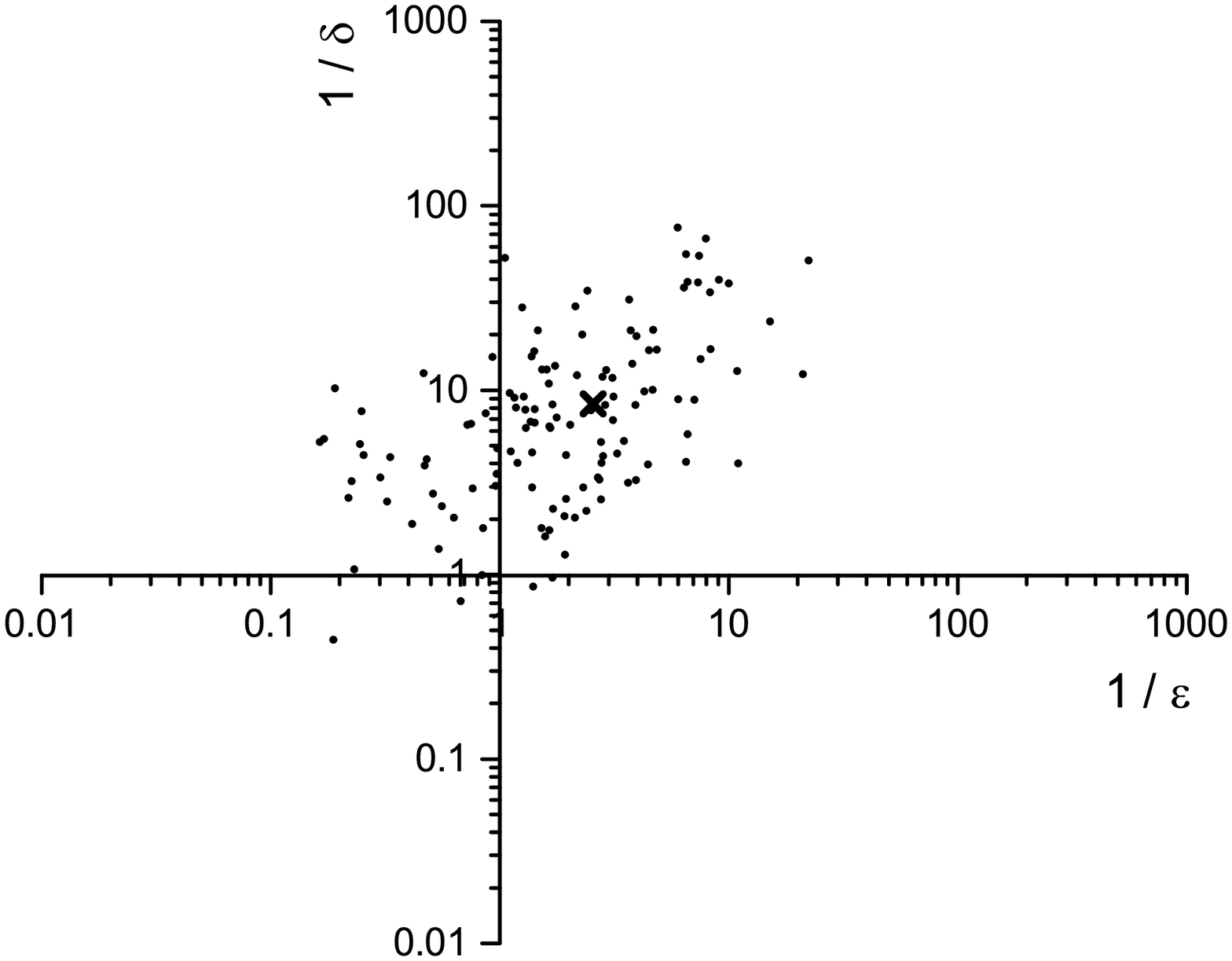}
\end{center}
\end{figure}

\end{document}